\documentclass[twocolumn,showpacs,preprintnumbers,amsmath,amssymb]{revtex4}  

\usepackage{graphicx}

\usepackage{dcolumn}

\usepackage{bm}

\begin{document}

\preprint{APS/123-QED}

\title{Soft gamma-ray background and light Dark Matter annihilation\\}

\author{Yann      Rasera}       
\affiliation{CEA Saclay, B\^at.~709, 91191 Gif-sur-Yvette, France}

\author{Romain         Teyssier}        \email{romain.teyssier@cea.fr}
\affiliation{CEA Saclay, B\^at.~709, 91191 Gif-sur-Yvette, France}

\author{Patrick   Sizun}  
\affiliation{CEA Saclay, B\^at.~709, 91191 Gif-sur-Yvette, France}

\author{Michel Cass\'e}  
\affiliation{CEA Saclay, B\^at.~709, 91191 Gif-sur-Yvette, France}
\affiliation{Institut d'Astrophysique de Paris, 98bis Bd Arago,
75014 Paris, France}

\author{Pierre Fayet}  
\affiliation{Laboratoire de Physique Th\'eorique de l'ENS,
UMR 8549 CNRS, 24 rue Lhomond, 75231 Paris Cedex 05, France}

\author{Bertrand Cordier}  
\affiliation{CEA Saclay,{} B\^at.~709, 91191 Gif-sur-Yvette, France}

\author{Jacques Paul}  
 \affiliation{CEA Saclay,{} B\^at.~709, 91191 Gif-sur-Yvette, France}

\date{March 9, 2006}

\begin{abstract}

The bulk of the extragalactic background  between 10 keV and 10 GeV is
likely to  be explained by the  emission of Seyfert  galaxies, type Ia
supernovae,  and  blazars.   However,  as  revealed  by  the  INTEGRAL
satellite, the bulge  of our galaxy is an intense source  of a 511 keV
gamma-ray  line,  indicating  the  production  of a  large  number  of
positrons that annihilate.   The origin of the latter  is debated, and
they  could be  produced, in  particular,  by the  ($S$- or  $P$-wave)
annihilation of  light Dark Matter particles into  $\,e^+e^-$.  In any
case, the cumulated  effect of similar sources at  all redshifts could
lead to a  new background of hard $X$-ray  and soft gamma-ray photons.
On the basis of the hierarchical model of galaxy formation, we compute
analytically the  SNIa contribution to  the background, and add  it to
Seyfert  and  blazars  emission  models.  Confronting  these  expected
contributions to  observation, we find that any  extra contribution to
this unresolved background around 511 keV should be lower than about 4 keV
cm$^{-2}\!$  s$^{-1}\!$ sr$^{-1}$. We  also estimate  analytically the
extragalactic background  due to Dark  Matter annihilation, increasing
the  accuracy  of the  earlier computations. Indeed, we take  into
account the large  positron escape fraction from low  mass dark matter
halos, unable  to confine a dense and  magnetized interstellar medium.
Our new  background estimate  turns out to  be one order  of magnitude
{\it lower\,} so that the  hypothesis of a light Dark Matter candidate
remains compatible  with the  observed extragalactic background  for a
wider range of particle masses and cross-sections.

\end{abstract}




\maketitle

\section{Introduction}

\begin{figure}  

  \centering            

  \includegraphics[width=\hsize]{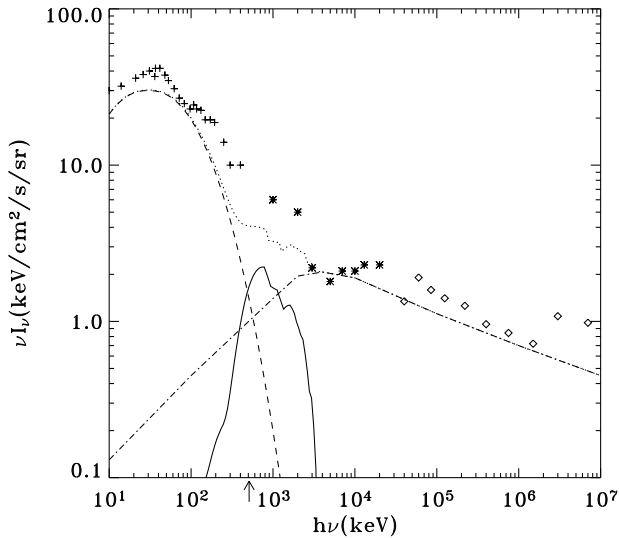}  

  \caption{Diffuse background spectrum as  a function of photon energy
  inspired by Fig.~4 of \citet{Strigari05} (we remove  the error bars
  for sake  of visibility).  The  crosses (HEAO), stars  (COMPTEL) and
  diamonds     (EGRET)      correspond     to     the     observations
  \citep{Zdziarski96,Strong04}.   At   low  energy,  Seyfert  galaxies
  (dashed  line) are  the main  contributors  \citep{Zdziarski96}.  At
  intermediate  energy,  Type  Ia  supernovae  (continuous  line),  as
  calculated  in  this article,  dominate.   At  high energy,  blazars
  (dot-dashed  line)  explain   the  observed  cosmological  gamma-ray
  background  \citep{Comastri99}.   Altogether the  sum  of the  three
  contributions (dotted line) is a  factor of 2 below the observations
  near 511~keV  (indicated by  an arrow).  An  additional contribution
  from light  Dark Matter  particles, of up  to about 4  keV cm$^{-2}$
  s$^{-1}$ sr$^{-1}$, \,is not excluded.}
  \label{contrib}  
 \end{figure}

The cosmic  gamma-ray background (CGB)  between $10$ keV and  $10$ GeV
has been measured by  several gamma-ray satellites (HEAO, SMM, COMPTEL
and  EGRET)  \citep{Zdziarski96,Strong04}.    Below  100  keV,  it  is
believed  that  the  main  contribution comes  from  Seyfert  galaxies
\citep{Zdziarski96}\,\footnote{It is  worth noting, in  addition, that
recent  deep  INTEGRAL  observations  \citep{Krivonos05} of  the  Coma
region suggest that  the cosmic ray background above  20 keV cannot be
explained in  terms of obscured  Seyfert galaxies.}.  Above 10  MeV, a
simple model for  blazars reproduces both the amplitude  and the slope
of  the data  \citep{Comastri99}.  In  the intermediate  energy range,
however, another type of sources  is needed, since blazar spectra show
a clear  break near 10  MeV and the cosmological  gamma-ray background
from   Seyfert  galaxies   falls  off   above  about   100   keV  (see
Fig.~\ref{contrib}).      As    discussed    by     several    authors
\citep{Watanabe99,Strigari05,Ruiz-Lapuente01},   type   Ia  supernovae
could make a  significant contribution in this energy  range, which we
shall evaluate in Section 2.

Furthermore, the  recent observation, by the INTEGRAL  satellite, of a 511  keV  diffuse    
emission    line     from    the    galactic bulge\,\citep{Knodlseder05} shows that electron-positron annihilation
are taking place there with  a very large rate $\simeq \,1.5\ 10^{43}$
s$^{-1}$.  Such a high rate is difficult, if not impossible, to impute to astrophysical  objects, and  the source of  these positrons  in the
bulge is subject  to intense debate.  This emission  from the bulge of the Milky  Way could be the  signature of light  Dark Matter particles annihilating  into   $e^+e^-$\  \cite{boehmfayet,fayet04},  the positrons  eventually annihilating with electrons  encountered in the interstellar   medium~\cite{boehmetal}.    In   any   case,   if   one extrapolates  this  diffuse emission  to  all  other  galaxies in  the Universe,  the integrated  flux  could make  a significant  additional
contribution to the hard $X$-ray and soft gamma-ray background.

  Since the  potential implications of  such an hypothesis  for particle physics and  cosmology are  very important, 
we  want to test  here its validity  or at  least its  consistency,  as far  as the  cosmological gamma-ray  background is concerned.   Using a  recent model  of galaxy formation~\citep{Rasera05}, we compute self-consistently the gamma-ray background    coming   from    both    type-Ia   supernovae    (adding Seyfert-galaxies and  blazars), and annihilating  positrons from light Dark Matter halos.  We follow in  this respect the earlier work of Ahn \&  Komatsu~\cite{Ahn05},   who  were   the  first  to   compute  this background, assuming, like them, however, that positrons  annihilate ``on the spot'' in all Dark  Matter halos  in the  Universe. These  authors  explored various scenarios for the internal structure of Dark Matter halos and analyzed how these  various models affect the amplitude of the gamma-ray background. 

In  the present paper, we  would  like to  go one  step further,  and explicitly take  into account  the role of  baryons in the  process of positron confinement and annihilation. This process is indeed possible only if  the parent halo  contains enough baryons (and  therefore also electrons) to host a dense, magnetized, interstellar medium. This is a necessary condition for the ``on the spot'' approximation to be valid. Since low mass  halos are unable to host enough baryons  in a cold and magnetized  disc~\citep{Rasera05}, the  escaping  positron mean  free path  increases dramatically.  To compute the positron  escape fraction  and their propagation  in the expanding background, a complex diffusion study would be necessary.   As a first order approximation we shall assume that  the positron escape fraction goes from zero to one below the critical mass for a Dark Matter halo to host a galaxy, as computed in~\citep{Rasera05}, and that these escaping positrons  never annihilate. 

The outline of  this article is as follows. In  Section 2, we estimate the contribution  of SNIa  to the gamma-ray  background, adding  it to that of  Seyfert galaxies and  blazars (at lower and  higher energies, respectively).  The difference with  the observed spectrum provides an upper  limit on any  additional contribution  such as  the one  due to annihilating  positrons,  that  could  come  from  light  Dark  Matter particle  annihilation.   In  Section  3, we  calculate  the  diffuse cosmological  background  induced by  all  Dark  Matter  halos in  the Universe,  taking into  account  that positrons  cannot annihilate  in small mass halos, and compare it to the previous calculation performed
by  Ahn  \&  Komatsu~\cite{Ahn05}.   In  Section  4,  we  present  the
gamma-ray background constraints on  the annihilation cross-sections and the masses of light Dark  Matter candidates.  We  summarize our main  conclusions in Section 5. In an Appendix, we estimate for calibration purpose, and confront with SPI/INTEGRAL observations, the 511 keV emission from the galactic bulge. We consider various annihilation cross-sections (depending on whether they are $S$- or $P$- wave dominated) and the corresponding halo Dark Matter density profiles.

\section{Diffuse gamma-ray background from SNIa}  
\label{Diffuse} 

\begin{figure} 
\centering            
\includegraphics[width=\hsize]{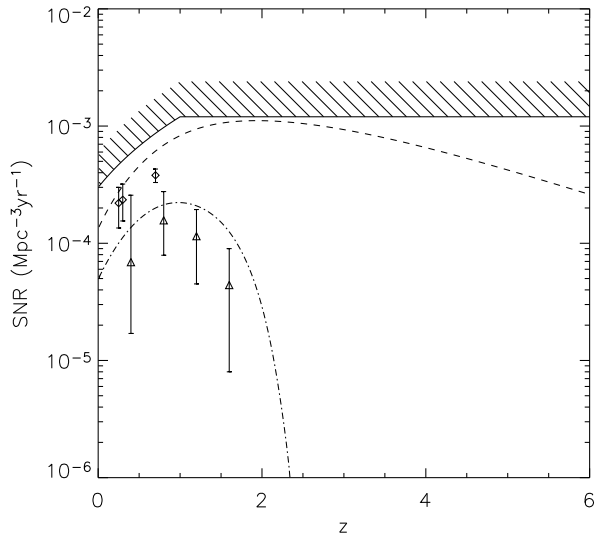}  
\caption{Comoving supernova rates as  a function of redshift from our analytical model. The  dashed line shows the SNII  rates and corresponds to the  star formation rate multiplied by  the fraction of SNII  per  unit  of   stellar  mass  formed  ($\epsilon_{SNII}  \simeq 0.007$~M$_\odot^{-1}$). The dot-dashed line represents the SNIa rate used for
computing the diffuse gamma-ray  background. Both rates are compatible
with the observations  from \citet{Dahlen04} (symbols) and  with the upper
limit from \citet{Strigari05} (dashed region).}
\label{snr}  
\end{figure}

The  type  Ia  supernovae  contribution to  the  gamma-ray  background depends  primarily on  the  star formation  history  in the  Universe, which, in this  paper, is derived from a  new self-consistent model of galaxy formation~\cite{Rasera05}.  This  analytical model predicts the cosmological evolution of the four main baryon phases in the Universe: diffuse intergalactic  gas, hot gas,  cold gas in galaxies  and stars. These  theoretical  predictions  were validated  with  high-resolution cosmological  simulations  using  the  RAMSES  \citep{Teyssier02}  and GADGET  codes \citep{Springel03b}.  They  also reproduce  the observed amount   of    cold   gas   in   the    Damped   Lyman-Alpha   systems \citep{Somerville01,Pei99} and the observed Cosmic Star Formation Rate \citep{Hughes98,    Steidel99,    Flores99,    Glazebrook99,    Yan99, Massarotti01, Giavalisco03,  Dahlen04} (see Fig.~\ref{snr}),  which is of  prime  interest  here.   The  model  particularly  emphasizes  the important cosmological  role of  the minimal mass  for a halo  to host galaxies, $M_{\rm min}(z)$. 

We shall use it to, first, compute the SNIa gamma-ray background, and, also, to evaluate  the new background that could  be attributed to the annihilation  of positrons, possibly  generated in  annihilating Dark Matter halos.  This internal consistency  allows us to perform  a fair comparison between the two types of gamma-ray sources. 

In the general case, the background intensity $I_\nu$ is given by
\begin{equation}  
I_{\nu}\ =\ \frac{c}{4\pi}\ \int^{t_H}_0  j_{\,\nu}(\,\nu(1+z)\,,\,z\,)\ \,dt\ ,  
\label{sum}  
\end{equation}  
with $j_{\nu}(\nu,z)$ the comoving emissivity at redshift $z$, or time
$t$.

The gamma-ray lines  from SNIa result from the  explosive synthesis of
radioactive $^{56}$Ni nuclei, decaying successively into $^{56}$Co and
$^{56}$Fe.  The comoving emissivity from SNIa may be expressed as
\begin{equation}  
j_{\nu}(\nu,z)\ \simeq\ h\nu\ \dot{\rho}_*(t-t_{SN})\ \epsilon_{SN}\   
\frac{M_{ej}}{A_{\rm Ni}\,m_p}  
\ S_{SN}(\nu)\ ,  
\end{equation}  
with $\ \dot{\rho}_*(t-t_{SN})\,$ the  comoving star formation rate at
time  $t-t_{SN}$ from  our analytical  model (see Fig.~\ref{snr}), \,$t_{SN}$  denoting the
average     delay     between      star     formation     and     SNIa
explosion. $\epsilon_{SN}$ is  the number of SNIa per  unit of stellar
mass   formed,  $M_{\rm  ej}$   the  mass   of  Nickel   produced  and
$S_{SN}(\nu)$  the  average  spectrum  per Nickel  nucleus.   We  take
$\,t_{SN}\simeq                       2.5\,$                      Gyr,
$\epsilon_{SN}=1.4\times10^{-3}$~M$_\odot^{-1}$ (so that the resulting
SNIa   rates  be  within   the  2$\sigma$   error  bars   of  observed
rates~\citep{Dahlen04}),  $M_{\rm ej}\simeq  0.5$~M$_{\odot}$,  and the
spectrum $S(\nu)$ as computed in \cite{Nomoto84}.

The resulting extragalactic background spectrum from SNIa is presented
in Fig.~\ref{contrib}.  It  shows a bump in the range 300  keV up to 3
MeV,  at  a  level which  turns  out  to  be  close to  the  predicted
contributions from Seyfert galaxies and blazars. This contribution, in
agreement  with  the  SNIa  contribution from  \citet{Strigari05},  is
slightly higher than their preferred model~\footnote{Our model is close
to the upper bound  of Fig. ~\ref{snr} in \citet{Strigari05}}, because
our star formation history is slightly more efficient, as suggested by
recent observations~\citep{Rasera05}.

Altogether the  resulting evaluation from  known astrophysical sources
reproduces reasonably well the observed extragalactic background below
$100$~keV and above  $3$~MeV. However, in the range  from $100~$keV to
$3$~MeV the  three contributions  fall short of explaining the bulk  of the
Cosmic  Gamma-Ray background  (as  emphasized by  \citet{Strigari05}).
Particularly,  in the  $100$~keV-$511$~keV  range of  interest in  this
article, the sum of the three contributions appears to be lower with a
difference of the  order of 4 keV cm$^{-2}$  s$^{-1}$ sr$^{-1}$.  This
sets an upper limit on a possible Dark Matter annihilation signal.

\section{Diffuse background from cosmological halos}

\subsection{Diffuse background}

The diffuse background is  simply the sum of
the   redshifted  emissions   from  positron   annihilation   in  all
cosmological  halos, in  principle at  all  redshifts (Eq.~\ref{sum}).
The comoving emissivity  can be computed by summing  up the individual
halo emissivities
\begin{equation}
\label{comoving}  
j_{\nu}(\nu,z)\ = \,\int_{M_{\rm min}}^\infty \!M\ \frac{dN}{d\,\ln M \,dV}\  
\frac{L_{\nu}(M,z)}{M}\ \,d\,\ln M\,  
\end{equation}  
where  $M_{\rm min}$  is  the  minimal mass  for  emitting halos.  The
luminosity per halo integrated up  to the radius of the halo $R_{200}$
is
\begin{eqnarray}  
L_{\nu}(M,z)&=&\int^{R_{200}}_0 P_{\nu}(r)\ 4 \pi \,r^2 \,dr\ ,  
\end{eqnarray}  
and $N(M,z)$ is  the Press-Schechter~\citep{Press74} mass function for
cosmological  halos.   Considering   positron  annihilation  ``on  the
spot'', the volume emissivity is given by
\begin{eqnarray}  
\label{emissivity}  
P_{\nu}(r)\ =\   
\frac{1}{2}\ \,S_{\rm pos}(\nu)\ \,\rho^{\ 2}_X(r) \ \,  
\frac{\left< \sigma \,v_{\rm rel}\,(r) \right>}{m_X^{\ 2}}\ \,,  
\end{eqnarray}  
with $\rho_X(r)$ the Dark Matter mass density profile, $\left< \sigma
\,v_{\rm  rel}\,(r) \right>$ is  the annihilation  cross-section,
the  factor   $\frac{1}{2}$  being  present   only  in  the   case  of
self-conjugate   Dark  Matter   particles\,  \footnote{For   {\it  non
self-conjugate}  Dark  Matter  particles  Eq.~(\ref{emissivity})  gets
replaced by
\vspace{-2mm}  
$$  
\vspace{-2mm}  
\ \ \ \ P_{\nu}(r)\ =\ S_{\rm pos}(\nu)\ \ \rho_X(r) 
\ \rho_{\bar X} (r)\ \ \frac{\left< \sigma \,v_{\rm rel}\,(r)\right>}
{m_X^{\ 2}}\ \ ,  
$$  and one  can  generally  assume equal  densities  for Dark  Matter
particles  and  antiparticles,  so  that $\,\rho_X(r)  =\rho_{\bar  X}
(r)\,=\,\frac{1}{2}  \  \rho_{{\rm  tot}\,(X+\bar  X)}\, (r)\  $.   }.

Positronium  annihilation  introduces  a  specific  emission  spectrum
$S_{\rm pos}(\nu)$,  with 25\% of the  energy injected in  the 511 keV
line, and the remaining 75\% spread over a 3$\gamma$ continuum.

\subsection{Dark matter density profile}

The  mass distribution  in  each Dark  Matter  halo is  in fact  quite
uncertain. \citet{Ahn05} have explored a wide range of
halo density profile parameters.  In this paper, we restrict ourselves
to Dark Matter distribution parameters at face value, as suggested by
$N$- body simulations, based on the following general fitting formula
\begin{equation}
\rho_X(r)\propto \frac{1}{x^{\gamma} (1+x^{\alpha})^{\frac{\beta-\gamma}{\alpha}}}\ , 
\end{equation}
where  $x=r/r_{s}$  with  $r_s$   the  scaling  radius corresponding to  the
concentration  parameter $c=R_{200}/r_s$  (typically  between $4$  and
$40$  depending on halo  mass and  redshift). $\gamma$,  $\alpha$ and
$\beta$   control  the   slope  respectively   for   small  ($r<r_s$),
intermediate ($r\simeq r_s$) and large radii ($r>r_s$).

The  concentration parameter defines whether halos  are rather  peaked ($c
\simeq 40$) or shallow ($c \simeq 4$).  Here again the mean value as a
function of redshift  and halo mass is given by  a fit on cosmological
simulations \citep{Bullock01},
\begin{equation}
\label{c}
c=max(4,4 \frac{1+z_c}{1+z})
\end{equation}
with $z_c$ the collapse  redshift given by $M_*(z_c)=10^{-2} M$, $M_*$
being the  non-linear mass at a redshift  $z$. This formula  is valid
only for  halos greater than $\simeq  10^6$~M$_{\odot}$. The behaviour
of the concentration parameter  for smaller halo masses, unresolved by
numerical simulation, is totally unknown.

The slope at the center of  Dark Matter halos is thought to be between
$\alpha=1$ and  $\alpha=1.5$.  We therefore consider  two extreme dark
matter profiles as given by Navarro, Frenk and White~\citep{Navarro97}
($\alpha=1$,  $\beta=3$  and  $\gamma=1$)  and  Moore~\citep{Moore99}
($\alpha=1.5$, $\beta=3$  and $\gamma=1.5$). Note  however that both
density  profiles   saturate  at   very  low  radius   $R_{min}$  when
$n(R_{min})\left<  \sigma  \,v_{\rm  rel}\,(r) \right>=1/t_H$  due  to
self-annihilation ($n(R_{min})=\rho_X(R_{min})/m_X$ is the dark matter
numeric density and $t_H$ is the age of the universe).

\subsection{Annihilation cross-section}

The  relic  abundance  of  Dark  Matter  particles  depends  on  their
decoupling  temperature,  which is  a  fraction  of  their mass  (i.e.
$T_{F\!}= \frac{m_X}{x_F}$  \,with $\,x_F\simeq\,$ 16  to 20 depending
on  $m_X$)   \,and  is,  roughly,  inversely   proportional  to  their
annihilation       cross-section       $\,\left< \sigma\,v_{\rm
rel}/c \right>_F\,$ at freeze-out.  The values required for a correct
abundance,  corresponding to  $\,\Omega_{\rm dm}\simeq  23\ \%  $, are
then, for such light particles, of the order of a few (up to $\approx$
10) picobarns (corresponding to $\left<\sigma\,v _{\rm rel}\right>_F 
\approx 10^{-25}$  cm$^3$   s$^{-1}$)  \citep{boehmfayet,fayet04},
depending  on whether  they  are  self-conjugate or  not,  and on  the
possible  velocity-dependence of  their annihilation  cross-section at
freeze-out.

Such  values  are  in  any  case rather  large  compared  to  ordinary
weak-interaction  cross-sections, especially  when dealing  with light
particles. This  necessitates an unusual,  more powerful, annihilation
mechanism, that could result from the exchanges of a new light neutral
gauge boson $U$, or, in the case of spin-0 Dark Matter particles, from
the      exchanges      of      new      heavy      (e.g.      mirror)
fermions~\cite{boehmfayet,fayet04,boehmetal,Umirror}.

A  rather  large  annihilation  cross-section 
could lead  to an  excessive continuum  of gamma-ray  photons at
various energies  (depending on $m_X$).   Cross-sections which behave,
at  least  to  a  large  extent,  proportionally  to  $v^2$  ($P$-wave
annihilation),  may therefore be  preferred~\cite{boehmes}, especially
at  lower  $\,m_X$.   The  residual  annihilation  of 
Dark Matter particles  in bulges of spiral galaxies  or in ellipticals
would then  include a  suppression factor that  could be, in  the pure
$P$-wave        case,       as        strong        as       $\,v_{\rm
halo}^{\,2}/v_F^{\,2}\,\approx\,10^{-5}$.       \,Furthermore,     and
independently of the above  argument, lighter Dark Matter masses $m_X$
tend  to be  preferred,  to avoid  excessive  gamma-ray production  as
compared to $\,e^+$ production, in our galaxy~\cite{fayet04,beacom}.

We shall  therefore consider annihilation  cross-sections parametrized
as $\,\sigma \, v_{\rm  rel} \,\simeq\, a+b\,v^2$, with $\left< \sigma
\,v_{\rm  rel}  \right>_F \  \approx  10^{-25}$  cm$^3$ s$^{-1}$\,  at
freeze-out (for a self-conjugate particle  -- or twice this value, for
a   non  self-conjugate   one).   And   explore  in   particular,  for
low-velocity halo particles, the two extreme situations $\left< \sigma
v_{\rm rel}  \right>\ \simeq \,a$  \ ($S$-wave) and  $\,\simeq \,bv^2$
\,($P$-wave  annihilation).  The  resulting emission  profiles deduced
from  a   given  Dark  Matter  profile  $\rho_X$   are  computed  from
$\,\left< \sigma v_{\rm rel}  \right>\ \rho_X^{\  2}/m_X^{\   2}\,$  
(cf.   Eq.~\ref{emissivity} for  a
self-conjugate  particle~\footnote{For {\it  non  self-conjugate} Dark
Matter    particles,     the    $\,\frac{1}{2}\    \rho_X^{\,2}$    in
Eq.~\ref{emissivity}  gets replaced  by  $\,\rho_X\,\rho_{\bar X}$  or
simply  $\,\frac{1}{4}\ \rho_{{\rm tot}\,  (X+\bar X)}^{\  \,2}$ while
the  annihilation cross-section  gets  doubled, so  that the  expected
emissivity and resulting emission profile remain the same.}). 
See also the Appendix for further comments.

Note  that  for the  pure  $P$-wave  cross-section,  the emissivity  now
depends  on the  Dark  Matter 3D  velocity dispersion~\cite{asc}.   We
therefore  compute $\sigma^2_{3D}$  as  a function  of  the radius  by
solving  the  Jeans  equation  for  a NFW  or  Moore  potential.   The
resulting emission profiles turns out to be less peaked than for a pure
$S$-wave cross-section.

\subsection{Role of the baryons}

\begin{figure}[h!]  

  \centering  
  \includegraphics[width=\hsize]{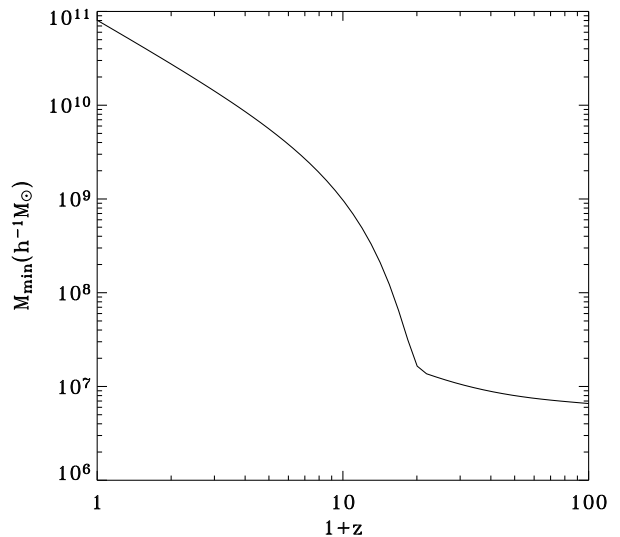}  
  \caption{Redshift evolution  of the minimal halo  mass $M_{\rm min}$
    below which  cold disk gas  cannot form (see  \citep{Rasera05} for
    details).  We use  a reionization  redshift $z_r  \simeq  \,20$ as
    suggested by WMAP.}
  \label{tf}  
\end{figure}

This rather  standard approach  has been applied  to compute  the soft
gamma-ray  background  in  \cite{Ahn05}, integrating  individual  halo
emissivity  over  the  Press   \&  Schechter  distribution  (given  by
Eq.~\ref{comoving}), using as lower  bound of the integration interval
the maximum between the Dark Matter free-streaming and the Dark Matter
Jeans masses.  This leads to a minimal mass ($M_{\rm min}$) equal to a
fraction  of solar  mass.   As a  consequence  the comoving  gamma-ray
emissivity  is dominated by  the cumulated  emission of  numerous small
mass  halos  (if  one  uses  the  concentration  parameters  given  by
Eq.~\ref{c}~\footnote{Moreover  the concentration parameters  of these
small  halos are  quite uncertain  because they  are smaller  than the
numerical resolution of cosmological simulations.}).

However, as we have already  discussed, small mass halos are unable to
retain gas  and annihilate positrons, and  therefore cannot contribute
to the  gamma-ray background in  which we are interested.   We  take 
into  account  the crucial  role  of  the baryons  (and
associated   electrons,  and   magnetic  fields)   in   confining  and
annihilating the Dark  Matter positrons. This trapping 
can be  achieved only if
both the density of the  interstellar medium and the galactic magnetic
field  are  sufficient.   Indeed  cosmological simulations  of  galaxy
formation in  the hierarchical  framework of structure  formation show
that    baryons     cannot    collapse    and     form    high-density
centrifugally-supported gas discs in halos  having a mass lower than a
minimal  value   $M_{\rm  min}  \,\approx\,10^7-10^{11}$   
h$^{-1}$ M$_\odot$.

This critical mass threshold is a key ingredient of the current galaxy
formation theory. Gnedin showed that the fraction of baryons decreases
strongly   in    halos   smaller   than    the   so-called   filtering
mass\,\citep{Gnedin00}, as  a consequence of  the non-zero temperature
of the  intergalactic medium which  prevents gas from  collapsing into
too small Dark Matter halos. \citet{Hoeft04} also showed that the halo
mass  must be  greater  than the  minimal  cooling mass;  if not,  the
fraction of baryons  is high but galaxies cannot  form because cooling
is  inefficient. The  resulting minimal  halo mass  $M_{\rm  min}$ for
galaxy formation is then the  maximum between the minimal cooling mass
and  the  filtering  mass,   as  computed  in  \citep{Rasera05}.   The
evolution   of  this   minimal  mass   with  redshift   is   shown  in
Fig.~\ref{tf}.   It is  of course  much larger  than the  one  used in
\citep{Ahn05}.

As computing accurately the escape fraction of positrons as a function
of halo  mass is  beyond the  scope of this  paper, we  shall consider
here,  for  simplicity,  that  below $M_{\rm  min}$,  essentially  all
positrons escape the  halos (and we neglect their  contribution to the
background), while above $M_{\rm  min}$, confinement is supposed to be
efficient and  all positrons are  taken to contribute.  Furthermore, the contribution
from  diffuse  baryons in  the  Universe  (baryons  which are  not  in
collapsed halos) is negligible in this range of wavelength because the
annihilation   time   scale   is   larger   than  the   age   of   the
Universe\,\footnote{Indeed,  direct annihilation of  positrons having
escaped out of small halos as well as annihilation through positronium
in the  intergalactic medium  are negligible (for  $z<50$) due  to the
very  low  density  of  the  latter  (about  $2  \  10^{-7}\  (1+z)^3$
cm$^{-3}$).}. Note that, while the ``on the spot'' approximation, used
by Ahn  \& Komatsu~\citep{Ahn05} for  the whole mass range,  leads, in
our  opinion, to  an  overestimation of  the  diffuse background,  our
approach, though  more accurate, should lead to  an underestimation of
the background level.

\begin{figure}[h!]  

  \centering  
  \includegraphics[width=\hsize]{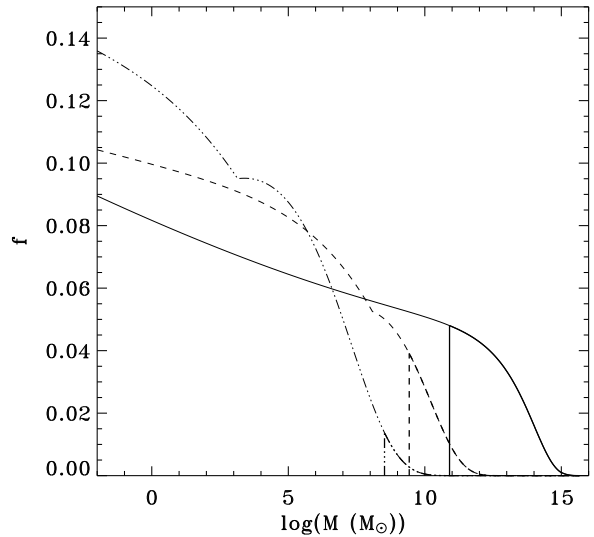}  
  \caption{Distribution of comoving  positron emissivity as a function
  of halo  masses for three different redshifts:  $z=0$ (continuous line),
  $z=6$ (dashed line) and $z=12$ (dot-dashed line). For each redshift,
  the vertical  line indicates the minimal  mass for a halo  to host a
  galaxy and  therefore to be able to annihilate positrons into  gamma-rays. The
  bold lines delineate the  regions, to the right of the vertical lines,
 where positrons are converted into photons. It follows that only about 10 \% of the 
 released positrons are actually converted into gamma-rays.}
  \label{npointfrac}  
\end{figure}

Fig.~\ref{npointfrac}   illustrates  the   distribution   of  comoving
positron  emissivity  as  a  function  of halo  masses  for  different
redshifts.   The  comoving  emissivity  is  dominated  by  small  mass
halos. However, only a small fraction of the halos contains enough gas
and magnetic field  for the production of gamma-rays:  altogether, only 
a fraction of about 10 \% of the emitted positrons are converted into
photons.  As a consequence, our
computation of the gamma-ray background leads to a result about 10 times smaller
than the one evaluated by \citet{Ahn05}. 

\begin{figure}[h!]  

  \centering  

  \includegraphics[width=\hsize]{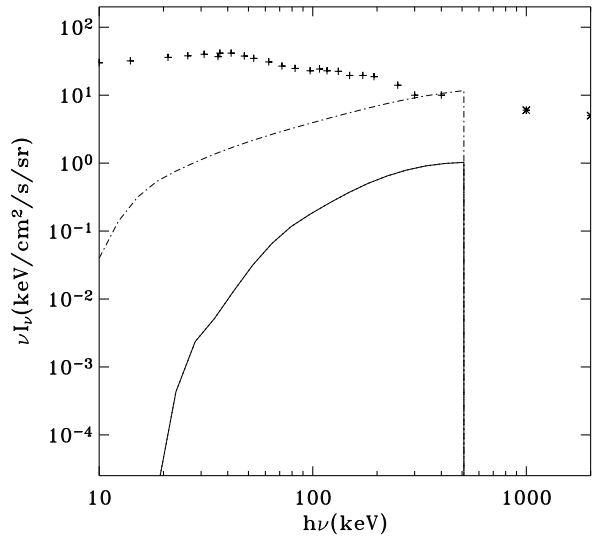}  

  \caption{Cosmic  gamma ray  spectrum produced  by NFW  profiles with $m_X=20$~MeV, a $S$-wave cross-section and, considering that no positronium is formed at all (for the purpose of comparison with earlier results). The upper dot-dashed curve  was computed  using the dark  matter-based minimal mass  (as  in  \citet{Ahn05}).   On  the  contrary,  for  the  lower continuous  curve we  took into  account baryon  physics,  introducing a minimal mass for a halo to  host a galaxy. The crosses and stars are the observational data.}
  \label{gamma_sigmav_ahn}  

\end{figure}

More precisely, we have
computed the soft gamma-ray  background for direct annihilation with a
$S$-wave  cross-section,  a NFW  Dark  Matter  density  profile and  a
particle  masse  $m_X=20$~MeV  in  the  two  different  cases  :  dark
matter-based minimal mass as in \citep{Ahn05} and baryon-based minimal
mass as in the present paper.  Fig.~\ref{gamma_sigmav_ahn} illustrates
the decrease  by a factor of 10  using our new approach.   Thus, if we
would  like to  reach the  same  level of background, we  would have  to divide  the
particle mass $m_X$ by a  factor of $\simeq \sqrt{10}$.  

Note that the
spectral shape is also modified.   The spectrum declines at low energy
because it corresponds  to high redshift where halos  greater than the
minimal  mass  for  galaxy  formation  becomes  rare.   Note  that  in
\citet{Ahn05},  identical results  were  recovered for  the same  Dark
Matter  particle  mass,  using  rather  extreme values  for  the  halo
concentration  parameter  $c$.   Recall  that here  we  consider  halo
concentration parameters only  at face value, as predicted  by $N$-body
simulations, but  we take into account  baryon physics as  part of the
annihilation mechanism.

\subsection{Calibration on Milky Way}

As  we have  seen  in  the previous sections,  the diffuse  gamma-ray
background depends  on three main  quantities.  The first is the annihilation cross-section: we  are going to explore two  extreme cases: $S$-wave and $P$-wave.   The second  ingredient is the  dark matter  mass density profile: we are going  to test peaked distributions (Moore, $c=15$) and shallow ones (NFW, $c=5$).   The last unknown quantity is the dark matter  particle  mass  $m_X$.   In  this Section,  we  are  going  to
fix the latter (for a given set of annihilation cross-section and density profile)
using the constraint set by the detected galactic signal.

The line  emission at 511 keV  detected by INTEGRAL  from the galactic
center  region is  at  a  level of  $10^{-3}$  ph cm$^{-2}$\  s$^{-1}$
\citep{Knodlseder05}.   Several types  of  astrophysical sources  have
been  considered as  potential  candidates to  explain this  emission.
However,  SNIa  fall  short  sustaining the  high  positron  injection
rate~\cite{Casse04,Prantzos04}.   Hypernovae~\citep{Schanne05} and the
related gamma-ray bursts~\citep{Casse04,Schanne05,Parizot05,Bertone04}
are in  a better position,  but since the  number of massive  stars is
about ten times larger in the disk than in the bulge, hot spots of 511
keV  emission should show  up in  the disk  plane\,\footnote{Except if
positrons  escape  rapidly  from  the  thin disk,  where  the  gas  is
concentrated.}, which is  not the case.  Low mass  X-ray binaries have
also been  suggested~\citep{Prantzos04}, but  no 511 keV  emission has
been observed from these objects. We therefore consider the hypothesis
that  light Dark  Matter particles  annihilate  into electron-positron
pairs,  mainly in  the galactic  center region  where the  Dark Matter
density is at a maximum\,\cite{boehmfayet,fayet04,boehmetal,asc}.  The
resulting low-energy  positrons are confined by the  magnetic field of
the  bulge, where  they are  progressively slowed  down  by ionisation
losses.  A  large  fraction  (0.93)  forms  positronium  with  ambient
electrons  and annihilate  into  two (25\%  of  probability) or  three
$\gamma$  photons  \citep{Guessoum05}.   Positronium  formation  plays
therefore an important role because it  decreases by a factor of about 3 the
intensity of the $511$~keV line.

The flux of gamma rays (from a direction making an angle $\theta$ with
the direction of the galactic center)  is given by the integral of the
emissivity along the line of sight
\begin{eqnarray}  
F_{\nu}(\theta)\ =\ \frac{1}{4\pi}\ \int_{\rm los} P_{\nu}(r)\  dl\ ,  
\end{eqnarray}  
where $P_{\nu}(r)$ is the volume  emissivity. We assume that essentially  
no  annihilation  can take  place
outside  the   stellar  bulge,   due to a lack of gas
(corresponding roughly to an angle $\theta_{\rm bulge}=16^\circ$).

The  resulting  profiles are  convolved  with  the INTEGRAL/SPI  Point
Spread Function (PSF). This  method allows  a fair  comparison between
observations and models,  and depends only weakly of  the poorly known
size of the gaseous  bulge.  For each couple of  cross-section - dark matter
profile,  we have  computed which  mass $m_X$  fits best  the observed
level of  511 keV emission.  The  results are summarized  in Table
\ref{tablecalib}.

\begin{table}
  \begin{tabular}{|cc|c|c|}
    \hline 
    $m_{\rm X}$ (MeV)&&$S$-Wave&$P$-wave\\
    \hline
    &c=15&1500&2.4  \\  
    Moore&c=10&900 &1.2  \\
    &c= 5&440 &0.44 \\
    \hline  
    &c=15&190 &0.42 \\
    NFW  &c=10&110 &0.20 \\
    &c= 5&45  &0.060\\
    \hline
  \end{tabular}
  \caption{This  Table  summarizes, for  different  cases, which  dark
    matter particle mass (in MeV)  is (or would be) required to reproduce the level of
    the INTEGRAL  signal from the  galactic bulge. We  explore different
    cross-sections ($S$-wave or $P$-wave),  different inner slopes (Moore or
    NFW) and different concentration parameters ($c=5-10-15$). (A result smaller than $\frac{1}{2}$ MeV 
indicates that the signal cannot be reproduced with the annihilation cross-section and density profile considered.)}
  \label{tablecalib}
\end{table}

As expected,  the most peaked  profiles  (such as  Moore  profile) and the
most concentrated  ones  ($c=15$)  require the  largest  values of $m_X$, since for a given 
mass density the
number density, and hence the annihilation rate decreases when the 
Dark Matter particle mass increases.
It is worth noting that there is a factor of
$10^2-10^3$  for the mass (corresponding to $10^4-10^6$  in flux)  in favor  of the
$S$-wave cross-section compared  to the $P$-wave one. Both  the NFW $S$-wave
case and the  Moore $P$-wave reproduce the total flux of  the bulge $511$ keV
emission with  reasonable Dark Matter  particle mass of the  order of
 $m_X \simeq  100$~MeV  and $m_X  \simeq  1$~MeV, respectively. On the
opposite,   the  NFW   $P$-wave   case  would require   masses
($m_X<0.42$~MeV) so small that they are  unable  to  produce  511  keV
photons.   And  the  Moore  $S$-wave  case  required such high  masses
($m_X>440$~MeV) that they would lead to an excessive  bremsstrahlung emission of soft gamma-ray photons
 \cite{beacom}.

Among the specific cases considered two models (NFW  with $S$-wave  and  Moore with  $P$-wave) 
are  therefore
favored  by the  galactic signal.   If  one considers  only the  total
emission  from   the  bulge,  neither   the  $S$-wave  nor   the  $P$-wave
cross-section can be excluded.  If one considers the {\it emission
profile} however,  a different  conclusion could  be drawn:  according to a  recent article
\cite{asc} the {\it  shape} of  the  emission profile
could  be  used  to  exclude  the $P$-wave  scenario. We  address  this 
interesting  question  in the Appendix, 
in  which we conclude  that, to our opinion,  both scenarios
cannot be discriminated yet.

\subsection{Results}

\begin{figure}[h!]  
  \centering  
  \includegraphics[width=\hsize]{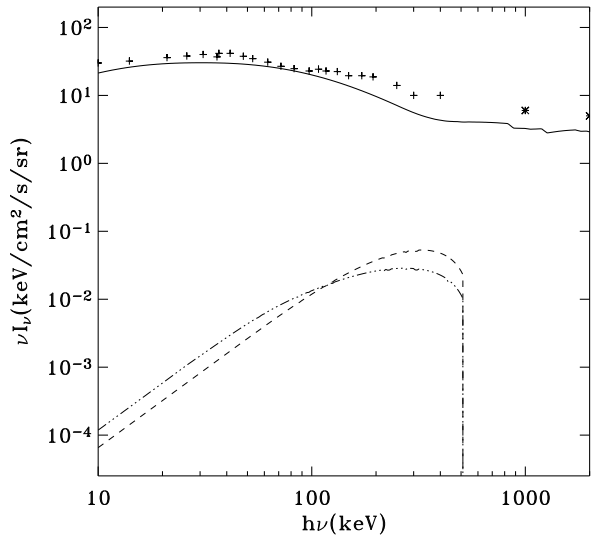}  
  \caption{Diffuse  background spectrum.   Crosses: HEAO  data; stars:
    COMPTEL  observations.  The  continuous  line is  the  sum of  the
    contributions    from   Seyferts,    SNIa   and    blazars.    The
    dot-dot-dot-dashed line and the dashed line  represent the revised  positron contribution
    to   the  cosmological  gamma-ray   background,  for   a  $S$-wave
    cross-section with a NFW profile, and a $P$-wave cross-section 
    associated to a  Moore profile, respectively.}
  \label{fondgamma}  
\end{figure}

Using the calibration on the Milky Way, we are now able to compute the
diffuse background  for our two  best models:
$S$-wave cross-section,  NFW profile with  a $100$~MeV Dark  Matter mass;
and  $P$-wave cross-section, Moore  profile with  a $1$~MeV  Dark Matter
mass. These are used here as specific benchmarks for the the purposes of our analysis, 
many other intermediate situations being obviously also possible.

As shown in  Fig.~\ref{fondgamma}, the background predictions in
the two models are at the  same level with a slightly different spectral
signature.  The  main conclusion  is that the  level of  the predicted
background  is  more   than  a  factor  of  100   below  the  observed
background. Calibrating  on the Milky  Way, the relative  smallness of
the  obtained results for the cosmic gamma-ray background shows  that the  light Dark  Matter annihilation
hypothesis   is  by far not   ruled-out   by  the   current  soft   gamma-ray
extragalactic background constraint.  

Note that this conclusion can be
applied more  generally to other positron sources  since the $511$~keV
emission from  the Milky Way  is quite weak  compared to  the observed
background intensity.  In order to make this background large, one has
to  assume that other  galaxies have  much higher  positron production
rates than the Milky Way. Indeed, in this case, the Milky Way would not be
representative from other halos of the same mass.

\section{Constraints on dark matter candidates}

\begin{figure}[h!]  
  \centering  
  \includegraphics[width=\hsize]{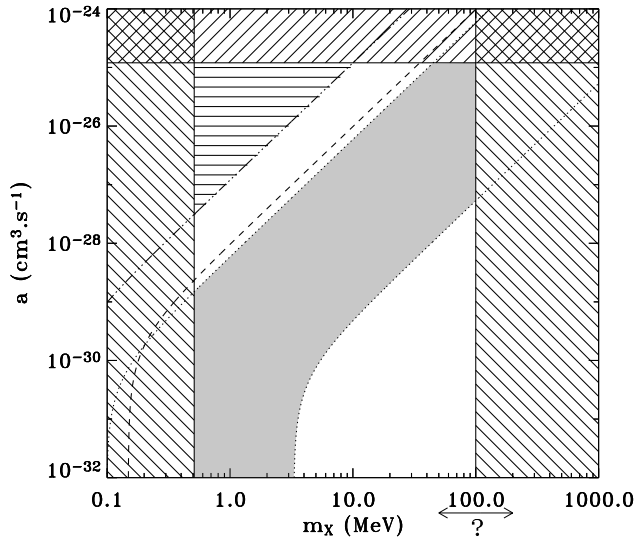}  
  \caption{Constraints  on the dark  matter candidate in the $a-m_X$
  plane, with  $a$ the velocity-independent part in the annihilation cross-section
$\,\sigma\,v_{\rm  rel}\,$ 
  and $m_X$ the  mass of the dark matter particle. 
$b$ is fixed by the relic density requirement, so that
$\left<\sigma\,v_{\rm  rel}   \right>_F \simeq \ <a+b\,v^2>_F\ 
  \approx   10^{-25}$  cm$^3$  s$^{-1}$\,   at  freeze-out. 
The   upper    line   $a\simeq 10^{-25}$~cm$^3$ s$^{-1}$
  corresponds  to  a  purely $S$-wave  cross-section, and
the lower  part of the diagram to a $P$-wave  dominated one ($a$ being
  negligible).  As $m_X$ cannot be too large, in order not to overproduce gamma and radio continuum from the galactic center, we limit ourselves, conservatively, 
to the interval 0.511 to 100 MeV (the actually-allowed mass interval could in fact be significantly smaller \citep{beacom,Casse05,Beacom05,Fayet06} depending on how seriously these other constraints are taken).\\
The grey region is the one compatible with the  galactic  constraint, based on the total level of emission, 
using the (Moore or NFW) dark-matter distributions of Table~I
(its upper part corresponds to a Milky-Way emission that would-be $S$-wave-dominated,
with $ a$ behaving like $1/m_X^{\ 2}$).\\
The dot-dashed  line is associated with a dark-matter
induced background (for a NFW  profile) that would correspond to the missing $4$
keV cm$^{-2}$  s$^{-1}$ sr$^{-1}$,
the top left-hand corner above this line being excluded on the basis of the cosmic background data (same for the dashed line with Moore profiles).
}
  \label{contraintes}
\end{figure}

The main objective of this paper is the computation of the Dark Matter
induced  background in the  100~keV-511~keV energy  range \emph{taking
into  account   the  important  role  of   baryons}.   Using  standard
concentration parameters  for the Dark Matter halo  profiles (as given
by N body simulations), the net  result of this exercice is a decrease
by a factor of 10 of  the level of the background emission compared to
the precedent computation in \citep{Ahn05}.

Calibration of $\left< \sigma  \,v_{\rm rel}\,(r) \right>$ and $m_X^{\
2}$ on  the INTEGRAL signal  is however uncertain. Indeed,  it depends
strongly  on the  Milky-Way Dark  Matter  profile which  is not  well
known.  The grey region  in Fig.~\ref{contraintes}  shows the  range of
allowed parameters corresponding to  this calibration.  To obtain this
domain we have  considered a reasonable range of  Dark Matter profiles
from peaked  and concentrated  ones (Moore profile  and $c=15$)  to less
peaked and  less concentrated ones (NFW  and $c=5$). Then  we have found
for each  profile which cross-section and particle  mass reproduce the
level  of  the  galactic  emission.    As  a  result,  each  mass  and
cross-section  in the  grey region  could reproduce  the level  of the
bulge  emission   with  a   reasonable  Dark  Matter   profile.   The
corresponding  background is  below the  observed background  (see the
previous Section).   As a consequence, no light  Dark Matter candidate
is  ruled out  by the  diffuse background  constraint and  one  has to
invoke  another gamma-ray  source in  order to  explain the  missing 4
keV~cm$^{-2}$~s$^{-1}$~sr$^{-1}$.

If we  consider now  that the  Milky-Way Dark  Matter halo  is totally
different from other  halos of similar mass in  the Universe, we could
relax the  previous calibration and obtain  independent constraints on
the mass and  the cross-section (following in that  sense the strategy
used in \cite{Ahn05} on the observed background only).  In the present
paper, constraints are of  course less stringent than in \cite{Ahn05},
since baryon physics has led us  to decrease the level of the emission
by a  factor of 10.   As shown in  Fig.~\ref{contraintes}, constraints
exclude  only the  upper left hand  corner, corresponding  to  low-mass Dark
Matter particles  and $S$-wave dominated  cross-section.  Interestingly,
for  halos  with a  Moore  profile,  the Dark  Matter candidate  mass  and
cross-section required  to reach the level of  the observed background
are not so far from the region favored by the galactic constraints.

\section{Conclusions}

Having estimated the SNIa contribution in the 100 keV-10 MeV energy range,
we  have found that  an unexplained  gamma-ray background  emission at
most of the  order of 4 keV cm$^{-2}$  s$^{-1}$ sr$^{-1}$ remains.  As
proposed in  \citep{boehmetal}, the strong  511 keV emission  from the
galactic center detected by  the INTEGRAL satellite could be explained
by  light Dark  Matter annihilation,  and we  have verified  that the
observed emission profile can be reproduced, both for $S$ and $P$-wave
annihilation   cross-sections.   Using   the  hierarchical   model  of
structure  formation,  we have  computed  the corresponding  gamma-ray
background, and  found it to be compatible  with current observational
bounds,  {\it if  one takes  into account  the minimal  halo  mass for
galaxy formation}.   The new positron-generated  (Dark Matter-induced)
extragalactic  background is  in fact  overwhelmed by  other emissions
from SNIa, Seyferts,  and blazars.  The exclusion of  small mass halos
as (redshifted) 511 keV photon  sources leads to an order of magnitude
decrease of  the extragalactic flux around 500 keV as
compared to earlier studies.   The spectral shape of the extragalactic
background  is also  modified  in  the sense  that  the number of Dark Matter  halos
capable  of hosting  gas  rich galaxies  decrease  very strongly  with
increasing redshift.

\vskip 2mm

\noindent

{\it Acknowledgements:} The authors  would like to thank the anonymous
referee  for  his helpful  remarks  that  have  greatly improved  the
quality of the paper.

\appendix

\section{511 keV emission from the galactic bulge}  

\label{mw}

The computation of  the Milky-Way emission profile  $F(\theta)$ is not
the  main  goal of  this  paper, since  we  only  used the  integrated
emission (and not the shape)  for calibration purposes. However, it is
essential to compare carefully predicted profiles to the observed one (see
the  recent paper  \cite{asc}). In  this  Appendix, we  would like  to
outline some  interesting issues concerning this point. Using the same hypothesis
as in Section  2,
we consider the  three different cases
of  Table \ref{tablecalib}:  a NFW  profile with  $c=10$ and  a $S$-wave
cross-section; a  NFW profile with $c=15$ and  a $P$-wave cross-section;
and finally, a Moore profile with $c=10$ and a $P$-wave cross-section.

\begin{figure}[h!]  
  \centering  
  \includegraphics[width=\hsize]{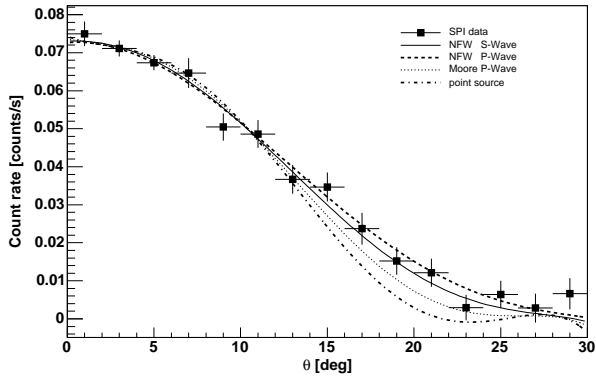}  
  \caption{ INTEGRAL/SPI overall counting rate  in the 511 keV line as
    a function of the angular distance to the galactic center.  Points
    with   error   bars   show   the   \emph{instrumental   background
    substracted} count  rate of the SPI  camera in the  506 to 516~keV
    band, averaged over the first year of
    \vspace{.5mm} data.   \break \hbox{\ \  \ \ \ }Lines  indicate the
    expected counting rates for three different situations:
    $S$-wave annihilation cross-section with a NFW Dark Matter profile
    (continuous line); \,or $P$-wave  cross-section with a NFW (dashed
    line) or  Moore (dotted line)  profile, as specified in  the text.
    The figure does not show the actual 511 keV emission profiles, but
    the  results of  their \emph{convolution  with  the spectrometer's
    response}, the Point Spread Function (PSF) -- which is a crude but
    more  rigourous way  in order  to compare  data with  models.  The
    lower  dot-dashed line  shows the  rate (not  compatible  with the
    data)  that  would correspond  to  a  pointlike  source.  For  the
    calibration, we choose  the Dark Matter particle mass  so that the
    observed integral of the signal over the inner 16 degrees is equal to
    the one from the theoretical profile.}
\label{fluxmwtot}  

\end{figure}

In   the   pure   $S$-wave   case   with   an   essentially   constant
\hbox{$\,\left<\sigma\,v_{\rm          rel}\,\right>\         \,\simeq
\,a\,\approx\,10^{-25} $ cm$^3$ s$^{-1}$},  we have chosen to consider
and test, with the above NFW distribution, a Dark Matter particle with
mass  $m_X =110$  MeV$/c^2$.   Let  us note  however  that, given  our
hypothesis, the same  emission profile would  have been obtained
from the same  $\,\rho_X$ but with a mass of  $\,\simeq 11$ \,(or 1.1)
MeV$/c^2$ only,  and an  $a$ term that  would be $\,\simeq  100$ \,(or
10$^4$) times  smaller.  The corresponding  ($S+P$-wave) cross-section
would then  be {\it$P$-wave dominated at  freeze-out}, while appearing
as {\it  $S$-wave-dominated\,} for low-velocity  annihilation {\it in
the galactic center}.

In  the  pure  $P$-wave  case   on  the  other  hand,  for  which  the
annihilation     cross-section      in     the     galactic     center
\,\hbox{$\left<\sigma\,v_{\rm   rel}\,\right>$}\,   should  be   lower
(i.e. typically $\,\approx 10^{-30}$ cm$^3$ s$^{-1}$), $m_X$ should in
general be  taken relatively  small, to get (with  a correct
relic density) a sufficiently  intense gamma-ray line.  In practice we
test $m_X$ about 0.5 (with the NFW profile) and 1.2 MeV$/c^2$ (with the Moore profile).

At first  sight, all  three tested profiles  seem compatible  with the
observations,  within the  precision of  the analysis.   Attempting to
discriminate  between  them (or  with  analogous  ones)  would require  a
careful chi-squared analysis  and is beyond the scope  of this article,
as we are  mainly interested here in the  total observed intensity and
global morphology of the 511 keV emission of the galactic bulge.

If  the  cross-section  for  Dark  Matter annihilation  in  halos  is
velocity-independent ($S$-wave  or effectively $S$-wave annihilation),
the emissivity  of a NFW  Dark Matter halo  scales near the  center as
$P_{\nu}\propto r^{-2}$.  Convolving this  emission with the SPI Point
Spread Function (PSF), we obtain the profile presented as a continuous
line in Fig.~\ref{fluxmwtot}. We emphasize here that the profiles shown
are the profiles after convolution by the PSF.

If, however, this cross-section is $S$-wave suppressed, the emissivity
now depends on the  Dark Matter 3D velocity dispersion~\cite{asc}.  We
then compute  $\sigma^2_{3D}$ as a  function of the radius  by solving
the  Jeans equation,  at first  for  a NFW  potential.  The  resulting
emission  profile  (shown  after  convolution  as  a  dashed  line  in
Fig.~\ref{fluxmwtot}) also turns  out to fit the data,  although it is
{\it less  peaked}, as  the velocities increase  with $r$,  within the
region of interest.  Furthermore, as mentioned earlier, to get in this
case  the  appropriate intensity  for  the 511  keV  line  we need  to
consider  both rather  small values  of $m_X$  (about $0.5$
MeV$/c^2$)  and somewhat  extreme parameters  for the  Milky -Way Dark
Matter halo (choosing $c\,\simeq\,15$).

These  restrictions may  be avoided to some extent,  however, with  a  {\it \,steeper
profile} such as the Moore profile~\citep{Moore99}, which enhances the
rate of Dark  Matter annihilation, especially near the  center of the
galaxy,  so   that  the   resulting  emission  profile   (shown  after
convolution as  a dotted line  in Fig.~\ref{fluxmwtot}) gets  now {\it
more peaked}. With such profiles, $P$-wave annihilation with standard
Milky-Way parameters  appear to be compatible with  the data, even for
less small  values of  $\,m_X$.  Furthermore, note  that {\it  if Dark
Matter  is  subject  to   the  stellar  gravitational  potential  that
dominates the  central region of the  galaxy}~\cite{robinetal}, with a
radial       density      profile      declining       roughly      as
$\,r^{-2}$~\cite{launhardtetal}, the $S$-  and $P$-wave cases would be
essentially indistinguishable.

To conclude this Appendix, we  have verified that one can reproduce the
photon flux and  distribution observed by INTEGRAL, both  for $S$- and
$P$ wave  cross-sections, with standard Dark  Matter profiles $\rho_X$
and appropriate mass $m_X$.  Again, in this  approach,  attempting to  
further  discriminate between  emission
profiles  associated with  $S$- or  $P$-wave annihilation  appears as
difficult, given the width of the PSF
function and  the variety of  the Dark Matter  profiles, gravitational
potential profiles  and gas density profiles which  may be considered.


\newpage  

  

\begin{thebibliography}{35} 

\expandafter\ifx\csname natexlab\endcsname\relax\def\natexlab#1{#1}\fi 

\expandafter\ifx\csname bibnamefont\endcsname\relax 

  \def\bibnamefont#1{#1}\fi 

\expandafter\ifx\csname bibfnamefont\endcsname\relax 

  \def\bibfnamefont#1{#1}\fi 

\expandafter\ifx\csname citenamefont\endcsname\relax 

  \def\citenamefont#1{#1}\fi 

\expandafter\ifx\csname url\endcsname\relax 

  \def\url#1{\texttt{#1}}\fi 

\expandafter\ifx\csname urlprefix\endcsname\relax\def\urlprefix{URL }\fi 

\providecommand{\bibinfo}[2]{#2} 

\providecommand{\eprint}[2][]{\url{#2}} 

 

\bibitem[{\citenamefont{{Zdziarski}}(1996)}]{Zdziarski96} 
\bibinfo{author}{\bibfnamefont{A.~A.} \bibnamefont{{Zdziarski}}}, 
  \bibinfo{journal}{MNRAS} \textbf{\bibinfo{volume}{281}}, \bibinfo{pages}{L9+} 
  (\bibinfo{year}{1996}). 

 

\bibitem[{\citenamefont{{Strong} et~al.}(2004)\citenamefont{{Strong}, 
  {Moskalenko}, and {Reimer}}}]{Strong04} 
\bibinfo{author}{\bibfnamefont{A.~W.} \bibnamefont{{Strong}}}, 
  \bibinfo{author}{\bibfnamefont{I.~V.} \bibnamefont{{Moskalenko}}}, 
  \bibnamefont{and} \bibinfo{author}{\bibfnamefont{O.}~\bibnamefont{{Reimer}}}, 
  \bibinfo{journal}{\apj} \textbf{\bibinfo{volume}{613}}, \bibinfo{pages}{956} 
  (\bibinfo{year}{2004}). 

 

\bibitem[{\citenamefont{{Krivonos et~al.}}(2005)\citenamefont{{Krivonos et~al.}, 
  {Vikhlinin}, {Churazov}, {Lutovinov}, {Molkov}, and {Sunyaev}}}]{Krivonos05} 
\bibinfo{author}{\bibfnamefont{R.}~\bibnamefont{{Krivonos {\em et~al.}}}}, 
  \bibinfo{journal}{\apj} \textbf{\bibinfo{volume}{625}}, \bibinfo{pages}{89} 
  (\bibinfo{year}{2005}). 

 

\bibitem[{\citenamefont{{Comastri}}(1999)}]{Comastri99} 
\bibinfo{author}{\bibfnamefont{A.}~\bibnamefont{{Comastri}}}, 
  \bibinfo{journal}{Astrophysical Letters Com.} 
  \textbf{\bibinfo{volume}{39}}, \bibinfo{pages}{181} (\bibinfo{year}{1999}). 

 

\bibitem[{\citenamefont{{Watanabe} et~al.}(1999)\citenamefont{{Watanabe}, 
  {Hartmann}, {Leising}, and {The}}}]{Watanabe99} 
\bibinfo{author}{\bibfnamefont{K.}~\bibnamefont{{Watanabe {\em et~al.}}}}, 
  \bibinfo{journal}{\apj} \textbf{\bibinfo{volume}{516}}, \bibinfo{pages}{285} 
  (\bibinfo{year}{1999}). 

 

\bibitem[{\citenamefont{{Strigari} et~al.}(2005)\citenamefont{{Strigari}, 
  {Beacom}, {Walker}, and {Zhang}}}]{Strigari05} 
\bibinfo{author}{\bibfnamefont{L.}~\bibnamefont{{Strigari {\em et~al.}}}}, 
  \bibinfo{journal}{astro-ph 0502150}  (\bibinfo{year}{2005}). 

 

\bibitem[{\citenamefont{{Ruiz-Lapuente} 
  et~al.}(2001)\citenamefont{{Ruiz-Lapuente}, {Cass{\' e}}, and 
  {Vangioni-Flam}}}]{Ruiz-Lapuente01} 
\bibinfo{author}{\bibfnamefont{P.}~\bibnamefont{{Ruiz-Lapuente}}}, 
  \bibinfo{author}{\bibfnamefont{M.}~\bibnamefont{{Cass{\' e}}}}, 
  \bibnamefont{and} 
  \bibinfo{author}{\bibfnamefont{E.}~\bibnamefont{{Vangioni-Flam}}}, 
  \bibinfo{journal}{\apj} \textbf{\bibinfo{volume}{549}}, \bibinfo{pages}{483} 
  (\bibinfo{year}{2001}). 

 

   

\bibitem{Knodlseder05}  
J. Kn\"odlseder \textit{et al.}, Astron. Astrophys. {\bf 411}, L457 (2003); 
P. Jean \textit {et al.}, Astron. Astrophys. {\bf 407}, L55 (2003);  
J. Kn\"odlseder \textit{et al.}, Astron. Astrophys. {\bf 441}, L513 (2003);  





   



















   


 

\bibitem{boehmfayet} C. Bo$\!e$hm and P. Fayet, Nucl. Phys. B {\bf 683}, 219 (2004).  
\bibitem[{\citenamefont{{Fayet}}(2004)}]{fayet04} 
\bibinfo{author}{\bibfnamefont{P.}~\bibnamefont{{Fayet}}}, 
  \bibinfo{journal}{\prd} \textbf{\bibinfo{volume}{70}}, 
  \bibinfo{pages}{023514} (\bibinfo{year}{2004}). 

 

\bibitem{boehmetal} C. Bo$\!e$hm {\textit {et al.}}, {\em Phys. Rev. Lett.} {\bf 92},  
 101301 (2004); C. Bo$\!e$hm, P. Fayet and J. Silk, Phys. Rev. D {\bf 69}, (2004) 101302 
 (2004). 

 

 

 

\bibitem[{\citenamefont{{Rasera} and {Teyssier}}(2005)}]{Rasera05} 

\bibinfo{author}{\bibfnamefont{Y.}~\bibnamefont{{Rasera}}} \bibnamefont{and} 
  \bibinfo{author}{\bibfnamefont{R.}~\bibnamefont{{Teyssier}}}, 
  \bibinfo{journal}{Astron. Astrophys.} \textbf{\bibinfo{volume}{385}}, \bibinfo{pages}{1}
  (\bibinfo{year}{2006}). 

 

\bibitem[{\citenamefont{{Ahn} and {Komatsu}}(2005)}]{Ahn05} 
\bibinfo{author}{\bibfnamefont{K.}~\bibnamefont{{Ahn}}} \bibnamefont{and} 
  \bibinfo{author}{\bibfnamefont{E.}~\bibnamefont{{Komatsu}}}, 
  \bibinfo{journal}{\prd} \textbf{\bibinfo{volume}{71}}, 
  \bibinfo{pages}{021303} (\bibinfo{year}{2005}); K. Ahn and E. Komatsu,  
Phys. Rev. D {\bf 72}, 061301 (2005); K. Ahn, E. Komatsu and P. H\"oflich, Phys. Rev. D {\bf 71}, 121301 (2005). 

 

 

\bibitem[{\citenamefont{{Teyssier}}(2002)}]{Teyssier02} 
\bibinfo{author}{\bibfnamefont{R.}~\bibnamefont{{Teyssier}}}, 
  \bibinfo{journal}{Astron. Astrophys.} \textbf{\bibinfo{volume}{385}}, \bibinfo{pages}{337} 
  (\bibinfo{year}{2002}). 

 

\bibitem[{\citenamefont{{Springel} and {Hernquist}}(2003)}]{Springel03b} 
\bibinfo{author}{\bibfnamefont{V.}~\bibnamefont{{Springel}}} \bibnamefont{and} 
  \bibinfo{author}{\bibfnamefont{L.}~\bibnamefont{{Hernquist}}}, 
  \bibinfo{journal}{MNRAS} \textbf{\bibinfo{volume}{339}}, \bibinfo{pages}{312} 
  (\bibinfo{year}{2003}). 

 

\bibitem[{\citenamefont{{Somerville} et~al.}(2001)\citenamefont{{Somerville}, 
  {Primack}, and {Faber}}}]{Somerville01} 
\bibinfo{author}{\bibfnamefont{R.~S.} \bibnamefont{{Somerville}}}, 
  \bibinfo{author}{\bibfnamefont{J.~R.} \bibnamefont{{Primack}}}, 
  \bibnamefont{and} \bibinfo{author}{\bibfnamefont{S.~M.} 
  \bibnamefont{{Faber}}}, \bibinfo{journal}{MNRAS} 
  \textbf{\bibinfo{volume}{320}}, \bibinfo{pages}{504} (\bibinfo{year}{2001}). 

 

\bibitem[{\citenamefont{{Pei} et~al.}(1999)\citenamefont{{Pei}, {Fall}, and 
  {Hauser}}}]{Pei99} 
\bibinfo{author}{\bibfnamefont{Y.~C.} \bibnamefont{{Pei}}}, 
  \bibinfo{author}{\bibfnamefont{S.~M.} \bibnamefont{{Fall}}}, 
  \bibnamefont{and} \bibinfo{author}{\bibfnamefont{M.~G.} 
  \bibnamefont{{Hauser}}}, \bibinfo{journal}{\apj} 
  \textbf{\bibinfo{volume}{522}}, \bibinfo{pages}{604} (\bibinfo{year}{1999}). 

 

\bibitem[{\citenamefont{{Hughes} et~al.}(1998)\citenamefont{{Hughes}, 
  {Serjeant}, {Dunlop}, {Rowan-Robinson}, {Blain}, {Mann}, {Ivison}, {Peacock}, 
  {Efstathiou}, {Gear} et~al.}}]{Hughes98} 
\bibinfo{author}{\bibfnamefont{D.~H.} \bibnamefont{{Hughes}}}, 
  \bibnamefont{\emph{et~al.}}, \bibinfo{journal}{\nat} \textbf{\bibinfo{volume}{394}}, 
  \bibinfo{pages}{241} (\bibinfo{year}{1998}). 

 

\bibitem[{\citenamefont{{Steidel} et~al.}(1999)\citenamefont{{Steidel}, 
  {Adelberger}, {Giavalisco}, {Dickinson}, and {Pettini}}}]{Steidel99} 
\bibinfo{author}{\bibfnamefont{C.~C.} \bibnamefont{{Steidel}}}, 
  \bibnamefont{\emph{et~al.}}, \bibinfo{journal}{\apj} \textbf{\bibinfo{volume}{519}}, \bibinfo{pages}{1} 
  (\bibinfo{year}{1999}). 

 

\bibitem[{\citenamefont{{Flores} et~al.}(1999)\citenamefont{{Flores}, {Hammer}, 
  {Thuan}, {C{\' e}sarsky}, {Desert}, {Omont}, {Lilly}, {Eales}, {Crampton}, 
  and {Le F{\` e}vre}}}]{Flores99} 
\bibinfo{author}{\bibfnamefont{H.}~\bibnamefont{{Flores}}}, 
  \bibnamefont{\emph{et~al.}}, \bibinfo{journal}{\apj} \textbf{\bibinfo{volume}{517}}, 
  \bibinfo{pages}{148} (\bibinfo{year}{1999}). 

 

\bibitem[{\citenamefont{{Glazebrook} et~al.}(1999)\citenamefont{{Glazebrook}, 
  {Blake}, {Economou}, {Lilly}, and {Colless}}}]{Glazebrook99} 
\bibinfo{author}{\bibfnamefont{K.}~\bibnamefont{{Glazebrook}}},  
  \bibnamefont{\emph{et~al.}}, \bibinfo{journal}{MNRAS} \textbf{\bibinfo{volume}{306}}, \bibinfo{pages}{843} 
  (\bibinfo{year}{1999}). 

 

\bibitem[{\citenamefont{{Yan} et~al.}(1999)\citenamefont{{Yan}, {McCarthy}, 
  {Freudling}, {Teplitz}, {Malumuth}, {Weymann}, and {Malkan}}}]{Yan99} 
\bibinfo{author}{\bibfnamefont{L.}~\bibnamefont{{Yan}}}, 
  \bibnamefont{\emph{et~al.}}, \bibinfo{journal}{\apj} 
  \textbf{\bibinfo{volume}{519}}, \bibinfo{pages}{L47} (\bibinfo{year}{1999}). 

 

\bibitem[{\citenamefont{{Massarotti} et~al.}(2001)\citenamefont{{Massarotti}, 
  {Iovino}, and {Buzzoni}}}]{Massarotti01} 
\bibinfo{author}{\bibfnamefont{M.}~\bibnamefont{{Massarotti}}}, 
  \bibinfo{author}{\bibfnamefont{A.}~\bibnamefont{{Iovino}}}, \bibnamefont{and} 
  \bibinfo{author}{\bibfnamefont{A.}~\bibnamefont{{Buzzoni}}}, 
  \bibinfo{journal}{\apj} \textbf{\bibinfo{volume}{559}}, \bibinfo{pages}{L105} 
  (\bibinfo{year}{2001}). 

 

\bibitem[{\citenamefont{{Giavalisco} et~al.}(2004)\citenamefont{{Giavalisco}, 
  {Dickinson}, {Ferguson}, {Ravindranath}, {Kretchmer}, {Moustakas}, {Madau}, 
  {Fall}, {Gardner}, {Livio} et~al.}}]{Giavalisco03} 
\bibinfo{author}{\bibfnamefont{M.}~\bibnamefont{{Giavalisco}}}, 
  \bibnamefont{\emph{et~al.}}, \bibinfo{journal}{\apj} \textbf{\bibinfo{volume}{600}}, 
  \bibinfo{pages}{L103} (\bibinfo{year}{2004}). 

 

\bibitem[{\citenamefont{{Dahlen} et~al.}(2004)\citenamefont{{Dahlen}, 
  {Strolger}, {Riess}, {Mobasher}, {Chary}, {Conselice}, {Ferguson}, 
  {Fruchter}, {Giavalisco}, {Livio} et~al.}}]{Dahlen04} 
\bibinfo{author}{\bibfnamefont{T.}~\bibnamefont{{Dahlen}}}, 
  \bibnamefont{\emph{et~al.}}, \bibinfo{journal}{\apj} \textbf{\bibinfo{volume}{613}}, 
  \bibinfo{pages}{189} (\bibinfo{year}{2004}). 

 

\bibitem[{\citenamefont{{Nomoto} et~al.}(1984)\citenamefont{{Nomoto}, 
  {Thielemann}, and {Yokoi}}}]{Nomoto84} 
\bibinfo{author}{\bibfnamefont{K.}~\bibnamefont{{Nomoto}}}, 
  \bibinfo{author}{\bibfnamefont{F.-K.} \bibnamefont{{Thielemann}}}, 
  \bibnamefont{and} \bibinfo{author}{\bibfnamefont{K.}~\bibnamefont{{Yokoi}}}, 
  \bibinfo{journal}{\apj} \textbf{\bibinfo{volume}{286}}, \bibinfo{pages}{644} 
  (\bibinfo{year}{1984}). 

 
\bibitem[{\citenamefont{{Press} and {Schechter}}(1974)}]{Press74} 
\bibinfo{author}{\bibfnamefont{W.~H.} \bibnamefont{{Press}}} \bibnamefont{and} 
  \bibinfo{author}{\bibfnamefont{P.}~\bibnamefont{{Schechter}}}, 
  \bibinfo{journal}{\apj} \textbf{\bibinfo{volume}{187}}, \bibinfo{pages}{425} 
  (\bibinfo{year}{1974}). 

 
\bibitem[{\citenamefont{{Bullock} et~al.}(2001)\citenamefont{{Bullock}, 
  {Kolatt}, {Sigad}, {Somerville}, {Kravtsov}, {Klypin}, {Primack}, and 
  {Dekel}}}]{Bullock01} 
\bibinfo{author}{\bibfnamefont{J.~S.} \bibnamefont{{Bullock}}}, 
\bibnamefont{\emph{et~al.}}, \bibinfo{journal}{MNRAS} \textbf{\bibinfo{volume}{321}}, \bibinfo{pages}{559} 
  (\bibinfo{year}{2001}). 



\bibitem[{\citenamefont{{Navarro} et~al.}(1997)\citenamefont{{Navarro}, 
  {Frenk}, and {White}}}]{Navarro97} 
\bibinfo{author}{\bibfnamefont{J.~F.} \bibnamefont{{Navarro}}}, 
  \bibinfo{author}{\bibfnamefont{C.~S.} \bibnamefont{{Frenk}}}, 
  \bibnamefont{and} \bibinfo{author}{\bibfnamefont{S.~D.~M.} 
  \bibnamefont{{White}}}, \bibinfo{journal}{\apj} 
  \textbf{\bibinfo{volume}{490}}, \bibinfo{pages}{493} (\bibinfo{year}{1997}). 

 

\bibitem[{\citenamefont{{Moore} et~al.}(1999)\citenamefont{{Moore}, {Quinn}, 
  {Governato}, {Stadel}, and {Lake}}}]{Moore99} 
\bibinfo{author}{\bibfnamefont{B.}~\bibnamefont{{Moore}}}, 
 \bibnamefont{\emph{et~al.}}, \bibinfo{journal}{MNRAS} \textbf{\bibinfo{volume}{310}}, 
  \bibinfo{pages}{1147} (\bibinfo{year}{1999}). 

 
\bibitem{Umirror} P. Fayet, Nucl. Phys. B {\bf 187}, 184 (1981); B {\bf 347}, 743 (1990); Phys. Lett. B {\bf 142}, 263 (1984). 


\bibitem{boehmes} C. Bo$\!e$hm, T. Ensslin and J. Silk, J. Phys. G {\bf 30}, 
279 (2004). 

 


\bibitem{beacom} J. Beacom, N. Bell and  G. Bertone, 
    Phys. Rev. Lett. {\bf 94}, 171301 (2005).  

 

\bibitem{asc} C. Bo$\!e$hm and Y. Ascasibar, Phys. Rev. D {\bf 70} (2004) 115013;  
  Y. Ascasibar {\em et al.}, astro-ph/0507142 (2005).   


\bibitem[{\citenamefont{{Gnedin}}(2000)}]{Gnedin00} 
\bibinfo{author}{\bibfnamefont{N.~Y.} \bibnamefont{{Gnedin}}}, 
  \bibinfo{journal}{\apj} \textbf{\bibinfo{volume}{542}}, \bibinfo{pages}{535} 
  (\bibinfo{year}{2000}). 

 

\bibitem[{\citenamefont{{Hoeft} et~al.}(2004)\citenamefont{{Hoeft}, {Yepes}, 
  {Gottl{\" o}ber}, and {Springel}}}]{Hoeft04} 
\bibinfo{author}{\bibfnamefont{M.}~\bibnamefont{{Hoeft}}}, 
\bibnamefont{\emph{et~al.}}, \bibinfo{journal}{Baryons in Dark Matter Halos}  (\bibinfo{year}{2004}). 

 
\bibitem[{\citenamefont{{Cass{\' e}} et~al.}(2004)\citenamefont{{Cass{\' e}}, 
  {Cordier}, {Paul}, and {Schanne}}}]{Casse04} 
\bibinfo{author}{\bibfnamefont{M.}~\bibnamefont{{Cass{\' e}}}}, 
 \bibnamefont{\emph{et~al.}}, \bibinfo{journal}{\apj} \textbf{\bibinfo{volume}{602}}, \bibinfo{pages}{L17} 
  (\bibinfo{year}{2004}). 

 

\bibitem[{\citenamefont{{Prantzos}}(2004)}]{Prantzos04} 
\bibinfo{author}{\bibfnamefont{N.}~\bibnamefont{{Prantzos}}}, 
  \bibinfo{journal}{astro-ph/0404501}  (\bibinfo{year}{2004}). 

 

\bibitem[{\citenamefont{{Schanne} et~al.}(2005)\citenamefont{{Schanne}, {Cass{\'e}}, {Cordier}, and {Paul}}}]{Schanne05}
\bibinfo{author}{\bibfnamefont{S.}~\bibnamefont{{Schanne}}},
\bibnamefont{\emph{et~al.}}, \emph{\bibinfo{booktitle}{35th COSPAR Scientific Assembly}} 
  (\bibinfo{year}{2005}), pp. \bibinfo{pages}{2307-+}. 

 

\bibitem[{\citenamefont{{Parizot} et~al.}(2005)\citenamefont{{Parizot}, 
  {Cass{\' e}}, {Lehoucq}, and {Paul}}}]{Parizot05} 
\bibinfo{author}{\bibfnamefont{E.}~\bibnamefont{{Parizot}}}, 
 \bibnamefont{\emph{et~al.}}, \bibinfo{journal}{Astron. Astrophys.} \textbf{\bibinfo{volume}{432}}, \bibinfo{pages}{889} 
  (\bibinfo{year}{2005}). 

 

\bibitem{Bertone04}  
G. Bertone \textit{et al.}, astro-ph/0405005 (2004). 



     

\bibitem[{\citenamefont{{Guessoum} et~al.}(2005)\citenamefont{{Guessoum}, 
  {Jean}, and {Gillard}}}]{Guessoum05} 
\bibinfo{author}{\bibfnamefont{N.}~\bibnamefont{{Guessoum}}}, 
  \bibinfo{author}{\bibfnamefont{P.}~\bibnamefont{{Jean}}}, \bibnamefont{and} 
  \bibinfo{author}{\bibfnamefont{W.}~\bibnamefont{{Gillard}}}, 
  \bibinfo{journal}{Astron. Astrophys.} \textbf{\bibinfo{volume}{436}}, \bibinfo{pages}{171} 
  (\bibinfo{year}{2005}). 

 
   

\bibitem{robinetal} A.~C. Robin, \emph{et al.}, Astron. Astrophys. 409, 523 (2003) 

 

\bibitem{launhardtetal} R. Launhardt, R. Zylka, and P.~G. Mezger, Astron. Astrophys. 384, 112 (2002)

\bibitem{Casse05} M. Cass\'e and P. Fayet, astro-ph/0510490

\bibitem{Beacom05} J.F. Beacom  and H. Yuksel, astro-ph/0512411
 
\bibitem{Fayet06} P. Fayet, D. Hooper and G. Sigl, hep-ph/0602169

\end{thebibliography}
\end{document}